\documentclass[traditabstract,letter,longauth]{aa}
\usepackage{graphicx}
\usepackage{txfonts}
\begin{document}

\title{Mapping the interstellar medium in galaxies with Herschel\thanks{Herschel 
is an ESA space observatory with science 
instruments provided by European-led Principal 
Investigator consortia and with important participation from NASA.}/SPIRE}

\author{
S.A. Eales\inst{1}
\and M.W.L. Smith \inst{1}
\and C.D. Wilson \inst{16}
\and G.J. Bendo \inst{3}
\and L. Cortese \inst{1}
\and M. Pohlen \inst{1}
\and A. Boselli \inst{4}
\and H.L. Gomez \inst{1}
\and R. Auld \inst{1}
\and M. Baes \inst{2}
\and M.J. Barlow \inst{5}
\and J.J. Bock\inst{6}
\and M. Bradford \inst{6}
\and V. Buat \inst{4}
\and N. Castro-Rodr{\'\i}guez \inst{7}
\and P. Chanial \inst{8}
\and S. Charlot \inst{9}
\and L. Ciesla \inst{4}
\and D.L. Clements \inst{3}
\and A. Cooray \inst{10}
\and D. Cormier \inst{8}
\and J.I. Davies \inst{1}
\and E. Dwek \inst{11}
\and D. Elbaz \inst{8}
\and M. Galametz \inst{8}
\and F. Galliano \inst{8}
\and W.K. Gear \inst{1}
\and J. Glenn \inst{12}
\and M. Griffin \inst{1}
\and S. Hony \inst{8}
\and K.G. Isaak \inst{13}
\and L.R. Levenson \inst{6}
\and N. Lu \inst{7}
\and S. Madden \inst{8}
\and B. O'Halloran \inst{3}
\and K. Okumura \inst{8}
\and S. Oliver \inst{14}
\and M.J. Page\inst{15}
\and P. Panuzzo \inst{8}
\and A. Papageorgiou \inst{1}
\and T.J. Parkin \inst{16}
\and I. P{\'e}rez-Fournon \inst{7}
\and N. Rangwala \inst{12}
\and E.E. Rigby \inst{17}
\and H. Roussel \inst{9}
\and A. Rykala \inst{1}
\and N. Sacchi \inst{18}
\and M. Sauvage\inst{8}
\and B. Schulz\inst{19}
\and M.R.P. Schirm\inst{16}
\and L. Spinoglio \inst{18}
\and S. Srinivasan \inst{9}
\and J.A. Stevens \inst{20}
\and M. Symeonidis \inst{15}
\and M. Trichas \inst{3}
\and M. Vaccari \inst{21}
\and L. Vigroux \inst{9}
\and H. Wozniak \inst{22}
\and G.S. Wright \inst{23}
\and W.W. Zeilinger \inst{24}
          }

\institute{School of Physics and Astronomy, Cardiff University,
  The Parade, Cardiff, CF24 3AA, UK
\and
Sterrenkundig Observatorium, Universiteit Gent, Krijgslaan 281 S9,
B-9000 Gent, Belgium
\and
Astrophysics Group, Imperial College, Blackett Laboratory, Prince
Consort Road, London SW7 2AZ, UK
\and
Laboratoire d'Astrophysique de Marseille, UMR6110 CNRS, 38 rue F.
Joliot-Curie, F-13388 Marseille France
\and
Dept. of Physics and Astronomy, University College London,
Gower Street, London WC1E 6BT, UK
\and
Jet Propulsion Laboratory, Pasadena, CA 91109, California Institute of Technology, Pasadena,
CA 91125, USA
\and
Instituto de Astrof{\'\i}sica de Canarias (IAC) and Departmento de Astrof{\'\i}sica, Universidad de la Laguna (ULL), La Laguna, Tenerife, Spain 
\and
CEA, Laboratoire AIM, Irfu/SAp, Orme des Merisiers, F-91191, Gif-sur-Yvette, France
\and
Institut d'Astrophysique de Paris, UMR7095 CNRS, 98 bis Boulevard Arago, F-75014 Paris, France
\and
Dept. of Physics \& Astronomy, University of California, Irvine, CA 92697, USA
\and
Observational  Cosmology Lab, Code 665, NASA Goddard Space Flight
Center Greenbelt, MD 20771, USA
\and
Department of Astrophysical and Planetary Sciences, CASA CB-389,
University of Colorado, Boulder, CO 80309, US
\and
ESA Astrophysics Missions Division, ESTEC, PO Box 299, 2200 AG Noordwijk, The Netherlands
\and
Astronomy Centre, Department of Physics and Astronomy, University of
Sussex, UK
\and
Mullard Space Science Laboratory, University College London,
Holmbury St Mary, Dorking, Surrey RH5 6NT, UK
\and
Dept. of Physics \& Astronomy, McMaster University, Hamilton,
Ontario, L8S 4M1, Canada
\and
School of Physics \& Astronomy, University of Nottingham, University
Park, Nottingham NG7 2RD, UK
\and
Istituto di Fisica dello Spazio Interplanetario, INAF, Via del Fosso
del Cavaliere 100, I-00133 Roma, Italy
\and
Infrared Processing and Analysis Center, California Institute of
Technology, 770 South Wilson Av, Pasadena, CA 91125, USA
\and
Centre for Astrophysics Research, Science and Technology Research
Centre, University of Hertfordshire, Herts AL10 9AB, UK
\and
University of Padova, Department of Astronomy, Vicolo Osservatorio
3, I-35122 Padova, Italy
\and
Observatoire Astronomique de Strasbourg, UMR 7550 Universit\'{e} de
Strasbourg - CNRS, 11, rue de l'Universit\'{e}, F-67000 Strasbourg
\and
UK Astronomy Technology Center, Royal Observatory Edinburgh, Edinburgh, EH9 3HJ, UK
\and
Institut f\"{u}r Astronomie, Universität Wien, Türkenschanzstr. 17,
A-1180 Wien, Austria}

   \date{Submitted to A\&A Herschel Special Issue}

\abstract{The standard method of mapping the interstellar medium in a galaxy, by
observing the molecular gas in the CO 1-0 line and the atomic gas in the 21-cm
line, is largely limited with current telescopes to galaxies in the nearby universe.
In this letter, we use SPIRE observations of the galaxies M99 and M100 to explore
the alternative approach of mapping the interstellar medium using the continuum
emission from the dust.
We have compared the methods by measuring
the relationship between 
the star-formation rate and the
surface density of gas in the galaxies using both methods.
We find the two methods give relationships with a similar dispersion,
confirming
that observing the continuum emission from the dust is a promising method of
mapping the interstellar medium in galaxies.}

\titlerunning{Mapping the ISM with SPIRE}

\authorrunning{S.A. Eales et al.}


\maketitle

\section{Introduction}

Two of the key measurements of a galaxy that are needed for any investigation
of galactic evolution are the star formation rate and, because stars form out of
gas, the mass of interstellar material. The standard method of measuring the 
gas mass
is to use the 21-cm line to measure the mass of the
atomic gas and the CO 1-0 line to 
measure the mass of the molecular material. The
conversion between 21-cm flux and the mass of atomic gas is unambiguous, but
the conversion factor between the CO luminosity and the mass of
molecular gas (the notorious `X factor') is uncertain both because the CO 
1-0 line is optically thick
and because the CO molecule is only tracing the much
larger number of unseen hydrogen molecules (e.g. Bell, Viti \& Williams 2007).
Both techniques suffer
from the problem that with current telescopes it is very difficult to measure
gas masses for galaxies beyond a redshift of $\simeq$0.2, a major 
limitation now that the Herschel Space
Observatory (Pilbratt et al. 2010)
is detecting thousands of galaxies in the high-redshift universe (e.g. Eales et
al. 2010).

The idea of using the continuum emission from the dust as an alternative
way of mapping the
interstellar medium (ISM) has a long history, dating back at least to Hildebrand
(1983). More recently, Guelin et al. (1993, 1995) and Boselli, Lequeux \&
Gavazzi (2002) have attempted to determine the X-factor using the continuum
emission from the dust and the assumption that the gas-to-dust ratio is
the same in both the atomic and molecular phase of the ISM. 
The problem in attempting to measure the mass of the ISM
from the continuum emission of the dust is that the conversion factors
between the submillimetre flux and
the mass of dust and between the dust mass and the gas mass
are both poorly known.
These constants have traditionally been estimated from galactic objects, such as reflection
nebulae (e.g. Hildebrand 1983), but James et al. (2002) have 
argued that a better approach
is to estimate them from the galaxies themselves, which
is the approach we follow here.

In this letter, we have used both methods to map the ISM in two
nearby galaxies, M99 and M100, which were two of the first targets
to be observed in the Herschel Reference Survey (Boselli et al.
2010), a survey with SPIRE (Griffin et al. 2010)
of 323 galaxies in a volume-limited sample of the local universe with distances
between 15 and 25 Mpc. To assess the accuracy of
the methods,
we have used these maps to investigate the relationship between
the star-formation rate in these galaxies and the density of the
gas out of which the stars form.
This is usually paramaterised by a power-law relationship between the star-formation rate
per unit volume and the gas density, $\rho_{SFR} \propto \rho_{gas}^N$ (Schmidt
1959), which 
for a constant scale height will result in
a relationship between the star-formation rate per unit area of the disk and the
surface density of the ISM which is also a power law 
with the same index: $\sum_{SFR} \propto (\sum_{gas})^N$.
This relationship is of great interest since
different theoretical models predict values of $N$ between 0.75 and 2
(Bigiel et al. 2008 and references therein), but in this paper we are less interested in the value
of $N$ than in using the fact that there is undoubtedly some intrinsic
relationship
between $\sum_{SFR}$ and $\sum_{gas}$ to assess the
errors, both statistical and systematic, in both methods for estimating the
surface-density of the ISM.

\section{Mapping the ISM using emission from the dust}

The method of James et al. (2002) is based on the
assumption
that a constant fraction of metals in the ISM
is incorporated in dust grains.
The mass of the ISM in the
galaxy is then given
by:
\smallskip
$$
M_{hydrogen} = {S_{\nu} D^2 \over \kappa_{\nu} B_{\nu}(T) Z \epsilon
f}  \eqno(1)
$$
\smallskip
\noindent in which $f$ is the ratio of the mass of metals to the mass of hydrogen in
a gas with solar abundance,
$\epsilon$ is the ratio of the mass of metals
in the dust to the total mass of metals, $Z$ is the metallicity of the galaxy in units of
solar metallicity, and $\kappa_{\nu}$ is the mass-opacity
coefficient of the dust at the frequency at which the flux has been measured. 
If the frequency is low enough, the Planck function reduces to the Rayleigh-Jeans
approximation and the gas mass depends only linearly on the dust temperature.
The importance of SPIRE 
is that
its long-wavelength channels (350 and 500 $\mu$m) mean that it is much
more likely this approximation is true than for previous space observatories such
as IRAS, ISO and Spitzer. 

\begin{table}
\caption{Fits}             % title of Table
\label{table:1}      % is used to refer this table in the text
\centering                          % used for centering table
\begin{tabular}{l c c c c}        % centered columns (4 columns)
\hline\hline                 % inserts double horizontal lines
Method & Galaxy & N & x-residuals & y-residuals \\    % table heading
\hline                        % inserts single horizontal line
CO/HI & M99 & 1.46$\pm$0.13 &  0.077 & 0.111 \\
dust & M99 & 1.77$\pm$0.22 & 0.086 & 0.152 \\
CO/HI & M100 & 1.38$\pm$0.08 & 0.070 & 0.096 \\
dust & M100 & 1.77$\pm$0.10 & 0.050 & 0.088 \\
\hline                                   %inserts single line
\end{tabular}
\end{table}

James et al. (2002) used abundance measurements of the local ISM to estimate a value
for $\epsilon$ of 0.46.
We have used a value for $f$ of 0.019 and a solar metal abundance of
$\rm 12 + log_{10}({[O] \over [H]}) = 8.69$ (Asplund et al. 2009).
The dust temperature can be estimated from the spectral energy
distribution of the galaxy and $Z$ can be estimated from optical spectroscopy.  
The only unknowns are therefore $M_{hydrogen}$ and $\kappa_{\nu}$, and so in principle if
one has a single galaxy for which one knows the gas mass one can estimate $\kappa_{\nu}$.
On the assumption that the properties of the dust and the fraction of metals in the dust do
not vary from galaxy to galaxy, one can 
then use this method to estimate the mass of the interstellar
medium in a galaxy for which there is not a direct measurement of the gas mass.

\begin{figure*}
   \centering
   \includegraphics[width=16cm]{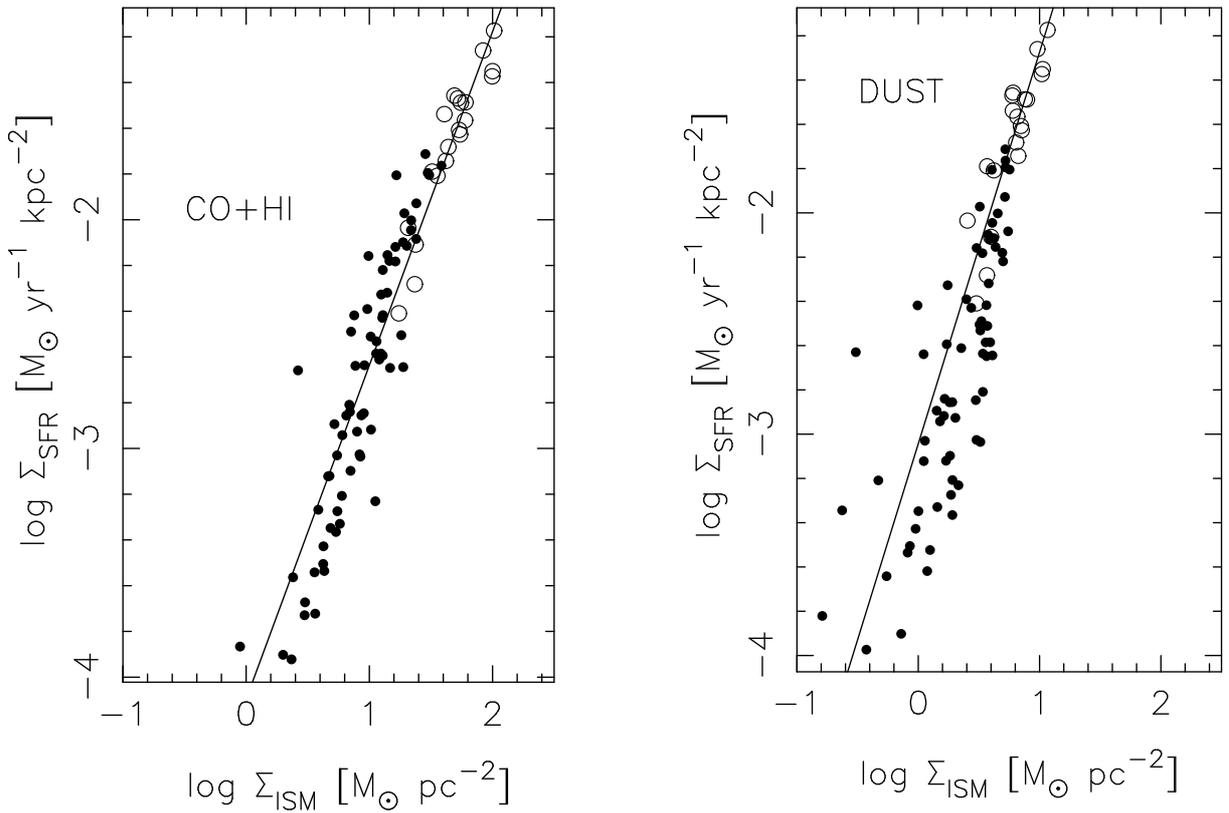}
      \caption{The star-formation rate per unit area of the galaxy verses the 
surface-density of the ISM for M99. In the left-hand panel the surface-density
of the ISM 
has been estimated from CO and HI measurements and in the right-hand panel
by the method described in the text.
The open circles show points within 80 arcsec of the centre of the galaxy, the region
in which
the surface-density of the molecular gas is greater than that of the atomic gas.
The line shows a fit to the points in this region.
} 
         \label{Fig. 1}
   \end{figure*}

\section{The Data}

M99 (NGC 4254) and M100 (NGC4321) are two 
spiral galaxies in the Virgo Cluster, which we take to have a distance
of 16.8 Mpc.
We have used the maps of the star-formation rate in these galaxies produced
by Wilson et al. (2009), who
combined 24-$\mu$m Spitzer images
and H$\alpha$ images to correct for the effect of dust obscuration, 
following the prescription of Calzetti et al. (2007).
We have produced maps of the gas in each galaxy by combining the
CO 1-0 maps of Kuno et al. (2007) and the HI maps from the VIVA survey (Chung et
al. 2009).
Before combining the maps, we convolved the CO map to the same resolution as
the 350-$\mu$m  map (FWHM of 25 arcsec). The HI map has slightly worse resolution 
than the 350-$\mu$m data and so we did not smooth it.
In estimating the mass of molecular 
material from the CO 1-0 line, we have used an X 
factor of $\rm 2 \times 10^{20}\ cm^{-2} (K\ km\ s^{-1})^{-1}$ (Bolatto et al. 2008).
In M99, the molecular material dominates the atomic gas within 80 arcsec of the galaxy centre
and in M100 within
100 arcsec. 

\begin{figure*}
   \centering
   \includegraphics[width=16cm]{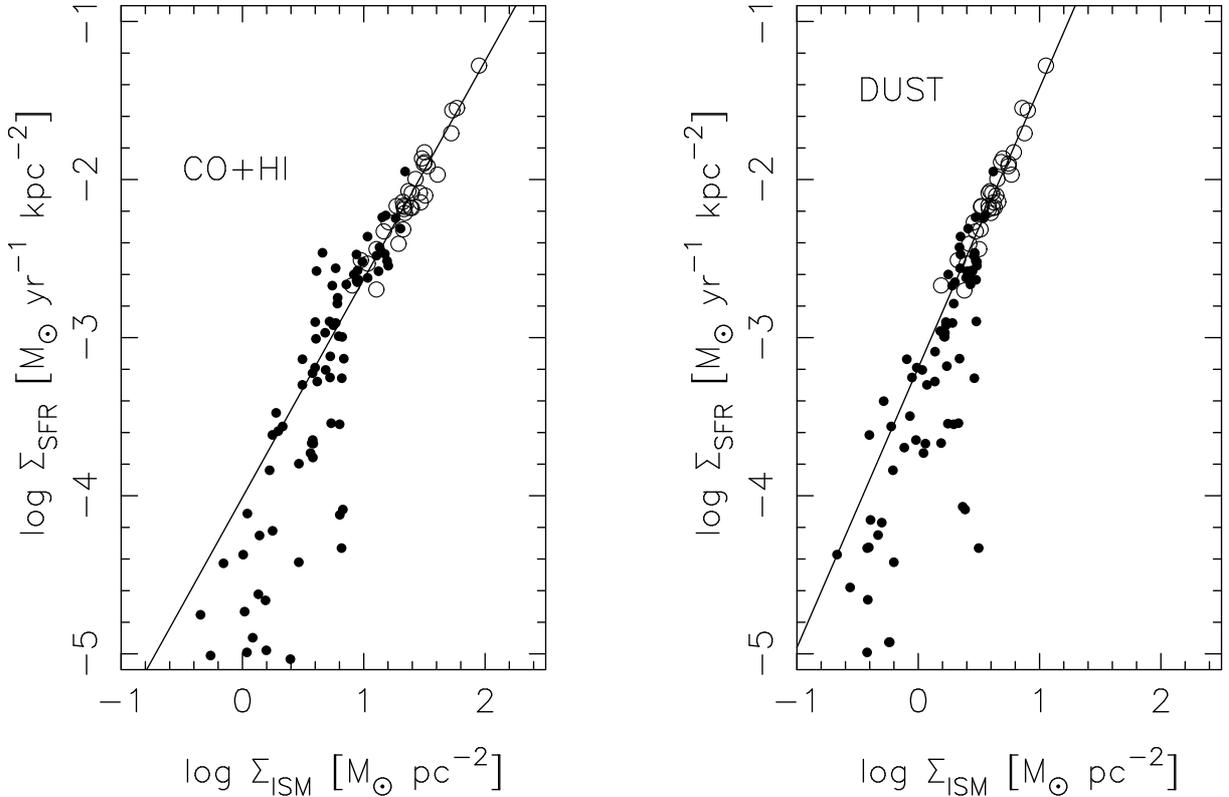}
      \caption{The same as Figure 1 but for M100. For this galaxy the
open circles show points within 100 arcsec of the centre of the galaxy, the region
in which
the surface-density of the molecular gas is greater than that of the atomic gas.
} 
         \label{Fig. 2}
   \end{figure*}

We observed M99 and M100 during the Herschel Science Demonstration Phase with SPIRE.
The calibration methods and accuracy of SPIRE are
described by Swinyward et al. (2010).
Images of the galaxies at the three SPIRE wavelengths (250, 350 and 500 $\mu$m)
are shown in Pohlen et al. (2010).
To estimate the distribution of dust temperature in each galaxy, which
is necessary to estimate the surface-density of the
ISM (Equation 1), we 
used the
SPIRE images at 250 and 350 $\mu$m
and the Spitzer images at 70 $\mu$m (the
SPIRE image at 500 $\mu$m does not have sufficient angular
resolution). We convolved the two short-wavelength maps to the
resolution of the 350-$\mu$m image and 
rebinned the images with a pixel size of 10 arcsec.
At each pixel, we fitted a single-temperature grey-body ($I_{\nu} \propto \nu^{\beta} B(\nu)$) 
with an emissivity index, $\beta$, of 2 to the three fluxes to estimate the 
dust temperature. 
The temperatures we derive are in the range $\rm 17 < T < 25$ K.
To determine the metallicity, we used the radial profiles
in Skillman et al. (1996).
The remaining unknown on the right of 
Equation (1) is $\kappa_{\nu}$. We estimated its
value at 350 $\mu$m by scaling, using $\beta=2$, from the value of 0.07 $\rm m^2\ kg^{-1}$ measured at
850 $\mu$m by James et al., giving a value
of 0.41 $\rm m^2\ kg^{-1}$. This is highly uncertain but,
as we will show below, it is possible in principle to recalibrate $\kappa_{\nu}$ using
the observations themselves. The flux density we used in Equation (1) was the
measured 350-$\mu$m flux density.
All the maps were rebinned into images with the same pixel size (10 arcsec) and
astrometrically registered.

\section{The Star-Formation Rates in M99 and M100}

Figs 1 and 2 show the star-formation rate per unit area plotted against the surface density
of the ISM for M99 (Fig. 1) and for M100 (Fig. 2). 
We have binned the images into 30-arcsec pixels before plotting these diagrams, so that 
the pixels are bigger than the beam size and
the data for
each point in the figures are clearly independent. At the distance of Virgo,
a 30-arcsec pixel corresponds to 2.4 kpc.
The left-hand panel in each figure shows
the result from the ISM map produced from the CO and HI data, and the right-hand panel shows
the result from using the ISM map produced from our alternative method.

The two panels look qualitatively very similar for both galaxies, with a tight relationship
in the region of each galaxy where the ISM is dominated by molecular material (the open circles)
and a much weaker relationship outside this region. There is a significant offset
between the two relationships along the abscissa,  
which must be due to systematic errors in one or both of the methods (\S 5).
For both methods and galaxies,
we used the bisector method (Isobe et al. 1990) to fit a straight line
to the points within the region dominated by molecular gas. 
On the assumption that there is an intrinsic relationship between the star-formation rate
and the surface-density of the ISM, one way to compare the accuracy of the two
methods of mapping the ISM is to compare the residuals around the best-fit lines.
Table 1 shows the 
root-mean-square residuals of the points around each
line along both the ordinate and the abcissa for both methods.
Based on this criterion, mapping the ISM from the dust emission works best
for M100 but mapping the ISM using the gas observations works best for
M99. 

There are, however, clearly systematic errors in one or both methods because the
values of $N$ determined from the two methods differ
by more than expected from the errors on each value. 
A simple test shows that this is not necessarily the result of a problem with the
dust method. We have repeated the analysis using
a value for the X-factor of
$\rm 1 \times 10^{20}\ cm^{-2} (K\ km\ s^{-1})^{-1}$, the value suggested
by Nakai and Kuno (1995). 
Restricting the analysis to the region in each galaxy in which the
molecular gas dominates, we find $N = 1.29\pm0.02$ for M99 and
$N = 1.67\pm0.07$ for M100.
The large changes show the different values of $N$ may be as easily
caused
by systematic errors in the X factor as
in our alternative method. 

Although the focus of this paper is on testing methods for mapping the
ISM in galaxies, we note that our results on the relationship between star-formation rate
and the surface density agree well with recent results.
Kennicutt et al. (2007) investigated this relationship within 
M51 and Bigiel et al. (2008) investigated it in 18 nearby
galaxies. Both studies concluded that there was little relationship between star-formation
rate and the surface-density of atomic gas but a strong one between the star-formation
rate and the surface-density of molecular gas. This is in qualitative  agreement with our finding
of a strong correlation in the inner part of each galaxy, where the molecular gas
is dominant. Bigiel et al. (2008) found that if one only considers the surface-density
of molecular gas the value of $N$ is $\rm \simeq 1$. Wilson et al. (2009) used CO 3-2 observations
to map the ISM in the same galaxies we have considered in this paper and found little variation
in the
gas depletion timescale across the disks, which implies $N\simeq1$.  
Our values of $N$ are higher than this, whichever method we use for
mapping the ISM. However, if we only use the CO 1-0 observations to map
the gas, we obtain $N = 1.14\pm0.10$ for M99 and $N = 1.02\pm0.08$ for M100, 
suggesting that it is the atomic phase of the ISM in the
central parts of these galaxies which is responsible for the higher values of $N$. 

\section{Calibrating the Mass-Opacity Coefficient}
 
We have attempted to calibrate $\kappa_{\nu}$ by requiring that the best-fit lines in the left-hand and right-hand panels of Figs 1 and
2 agree at $log \sum_{SFR} = -1.5$, a value chosen because it corresponds 
to the region in which the relationship
between
the gas surface-density and the star-formation density is tightest. 
We obtain a value for
$\kappa_{\nu}$ at 350 $\mu$m of 0.056 $\rm m^2\ kg^{-1}$ for M99 and 
0.063 $\rm m^2\ kg^{-1}$ for M100. 

These values are three times 
lower than the estimate from the theoretical models of Li and Draine (2001) 
of $\rm \simeq 0.19\ m^2\ kg^{-1}$. 
The only previous attempt to determine
the submillimetre opacity at this wavelength directly from observations
was that of Boulanger et al.
(1996), who used COBE observations of dust at high galactic latitude and
HI observations to obtain a relationship between submillimetre optical depth and
hydrogen column density.
Assuming solar metallicity and the same values of $f$ and $\epsilon$ used in this
paper (assumptions which are
equivalent to assuming a gas-to-dust ratio of 153 for the high-latitude gas), we derive a value
for $\kappa_{\nu}$ of 0.35 $\rm m^2\ kg^{-1}$, again higher than our value. Our
value is also, of course, significantly lower than the value (0.41 $\rm m^2\ kg^{-1}$)
obtained by extrapolating the measurement at 850 $\mu$m of James et al. (2002) using
a dust-emissivity index of 2.

The discrepancy between our result and the James et al. result is surprising, since the
method is the same in both cases, although James et al. estimated $\kappa_{\nu}$ from
integrated gas and dust measurements for a 
large sample of galaxies whereas we have estimated it from maps of the gas and dust
for indvidual galaxies. Since our result depends critically on the assumed value of the X factor,
one way to bring the results into line would be if the values of the X factor for M99 and
M100 are $\simeq$6 times lower than the value we have assumed, which seems
unlikely, however, given the recent evidence that the X factor is fairly constant
between galaxies (Bolatto et al. 2008). It seems more plausible that at least
part of the discrepancy is caused by our over-estimating the temperature of
the dust. When fitting single-temperature models, both non-equilibrium heating of grains
and the simple fact that warm dust produces more radiation than cold dust (Eales et al.
1989) can bias the fits towards higher temperatures. The temperatures produced
by our fits lie in the range $\rm 17 < T < 25\ K$ and the lack of any strong dependence
of dust temperature on star-formation rate agrees with the recent Herschel study
of the much closer galaxy M81 (Bendo et al. 2010). 
The temperatures are not 
much
higher than the 
temperature of high-latitude galactic dust measured by Boulanger et al. (1996): 17.5 K.
Nevertheless, if the temperature of the dust in both M99 and M100 were actually
$\simeq$10 K rather than 20 K, this would be enough to increase the value of $\kappa_{\nu}$ to
the value expected from the James et al. study.
There are other possible problems with this method, such as
disagreements between different methods
for measuring metallicity (Kewley \& Ellison
2008) and variations in $\kappa_{\nu}$ and $\epsilon$
both within and between galaxies, but
the biggest practical problem at present is the difficulty of accurately measuring the
temperature of the dust. Fortunately, this should be simple to overcome by
combining the Herschel observations with observations at longer wavelengths, for
example
at 850 $\mu$m with LABOCA and SCUBA-2.

In summary, the remarkable similarity between the right-hand and left-hand panels
of Figs 1 and 2 shows that observing the continuum emission from dust is
a promising method of mapping the interstellar medium in galaxies. Nevertheless,
it is clear that considerable work needs to be done to calibrate this method,
both in determining more accurately the temperature of the dust in individual galaxies
and in determing the variation in $\kappa_{\nu}$ and the
other properties of the dust in the galaxy population.
The Herschel Reference Survey and the other Herschel surveys of nearby galaxies
are ideal for this kind of study.

\acknowledgements{
SPIRE has been developed by a consortium of institutes led by
Cardiff Univ. (UK) and including Univ. Lethbridge (Canada);
NAOC (China); CEA, LAM (France); IFSI, Univ. Padua (Italy);
IAC (Spain); Stockholm Observatory (Sweden); Imperial College
London, RAL, UCL-MSSL, UKATC, Univ. Sussex (UK); Caltech, JPL,
NHSC, Univ. Colorado (USA). This development has been supported
by national funding agencies: CSA (Canada); NAOC (China); CEA,
CNES, CNRS (France); ASI (Italy); MCINN (Spain); Stockholm
Observatory (Sweden); STFC (UK); and NASA (USA).}

\end{document}